\documentclass[conference,a4paper]{APSIPA2021}
\usepackage{multirow}
\usepackage{graphicx}

\usepackage{amsmath}
\usepackage[psamsfonts]{amssymb}
\usepackage{amsxtra}
\usepackage{threeparttable}
\usepackage{multirow}
\usepackage{float}
\usepackage[colorlinks,linkcolor=blue]{hyperref}  
\usepackage[figure]{hypcap}
\begin{document}
\IEEEoverridecommandlockouts
\title{POSITIONAL-SPECTRAL-TEMPORAL ATTENTION IN 3D CONVOLUTIONAL NEURAL NETWORKS FOR EEG EMOTION RECOGNITION}
 
\author{%
\authorblockN{%
Jiyao Liu,
Yanxi Zhao, 
Hao Wu, 
Dongmei Jiang$^*$\thanks{$^*$Corresponding Author.} 
}
\authorblockA{
School of Computer Science and Engineering
\\ Northwestern Polytechnical University, Xi’an, Shaanxi 710072, China  \\
E-mail: 850143245@mail.nwpu.edu.cn 
}
 
}

\maketitle
\thispagestyle{empty}

\begin{abstract}
Recognizing the feelings of human beings plays a critical role in our daily communication. Neuroscience has demonstrated that different emotion states present different degrees of activation in different brain regions, EEG frequency bands and temporal stamps. In this paper, we propose a novel structure to explore the informative EEG features for emotion recognition. The proposed module, denoted by PST-Attention, consists of Positional, Spectral and Temporal Attention modules to explore more discriminative EEG features. Specifically, the Positional Attention module is to capture the activate regions stimulated by different emotions in the spatial dimension. The Spectral and Temporal Attention modules assign the weights of different frequency bands and temporal slices respectively. Our method is adaptive as well as efficient which can be fit into 3D Convolutional Neural Networks (3D-CNN) as a plug-in module. We conduct experiments on two real-world datasets. 3D-CNN combined with our module achieves promising results and demonstrate that the PST-Attention is able to capture stable patterns for emotion recognition from EEG. 

\end{abstract}
\section{Introduction}
Emotions play an important role in our daily life, including human communication, disease prediction, and so on. Even if emotions seem natural to us, we have little knowledge of the mechanisms behind the affective function of the brain [1]. EEG signal is capable of objectively reflecting different emotions. With the development of deep learning, research on emotion recognition based on EEG has attracted great interest in different interdisciplinary fields, including psychology detection, affective computing, and stress monitor [2,~3]. Although emotion recognition from EEG has achieved rapid development with machine learning, there are still many suspend problems to be solved and some negative factors in EEG signal to conduct emotion recognition [3,~4]. For example, the uncontrollable label tagging and the great differences in individual variations of emotional states make emotion recognition hard.
The EEG signal collected in the experimental environment is also influenced by many interference factors, e.g., time, context, space, races, language, culture [5], so there is much disturbing noise in the EEG signal. If we train the models based on the raw EEG samples indistinguishably, unnecessary noise will be introduced in the training processing. Therefore, the stability of EEG patterns, the validity of EEG features, and the rationality of model structure are vital to EEG emotion recognition and other real-world applications of EEG.

Although EEG is sensitive to differences in environment variables [7], it is considered to have some stable patterns exhibiting consistency among different sessions of the same participants or different participants in the same experimental setup [6]. Zheng et al. [6] investigate stable patterns of EEG over time for emotion recognition using a machine learning approach and show that power spectral density (PSD), differential entropy (DE), differential asymmetry (DASM), rational asymmetry (RASM), asymmetry (ASM) and differential caudality (DCAU) features are good indicators for EEG emotion recognition, which implies that taking different EEG features of different brain regions into consideration is beneficial to the task of emotion classification. Zheng et al. [8] show that using features on critical frequency bands and channels can get close or even better results than all EEG features without pruning. Wu et al. [9] indicate that combining the functional connectivity features from EEG and physiological signals can get promising results in emotion classification, which implies that brain regions' connectivity and topological structure are also key factors of emotion classification. On the other hand, as a time series, the time-variation of EEG signal is an important clue of emotion fluctuation. Although existing models have achieved high performance, most methods only use the raw EEG data or unprocessed EEG features and consider one or two factors of EEG emotion classification, but ignore the complementarity of the features among different dimensions, limiting the classification capability of the models to a certain extent. 

In this paper, we propose the Positional-Spectral-Temporal Attention (PST-Attention) in 3D convolutional neural networks (3D-CNN) to exploit important information in critical spectrum bands, brain regions, and time slots in the EEG signal for emotion classification. Specifically, the Positional Attention module learns the active regions induced by different emotions in the spatial dimension. The Spectral and Temporal Attention modules explore the different significance of frequency bands and temporal stamps, respectively. The main contributions of this paper can be summarized as follows: 

1)~We recombine the EEG features to a 4D shape as the EEG feature representation. It is an adaptive fusion of the position-spectral-temporal information of the EEG signal and a suitable input of our proposed structure. 

2)~We develop a parallel Positional-Spectral-Temporal Attention module to adaptively capture stable and discriminative patterns of spatial, spectral and temporal dimensions as well as suppress uninformative EEG features.

3)~Experiments are conducted on two benchmark datasets, and the results show that the 3D-CNN with PST-Attention Module outperforms most competitive methods.

The layout of the paper is as follows. In Section 2, we describe the related works on emotion recognition based on EEG. Section 3 presents the details of our proposed methods. Experimental evaluation and ablation experiments of the proposed methods are presented in section 4. Finally, we conclude our work in Section 5.

\section{RELATED WORK}
Emotion classification based on EEG signals is a meaningful research direction. As the electrophysiological manifestation of the central nervous system, EEG has the capability to reflect the real emotion states of humans objectively and precisely. With the development of artificial intelligence technology, emotion recognition has become a hot research topic in the field of human–computer interaction.

In recent years, traditional machine learning and deep learning methods have been widely used in EEG emotion recognition and other related applications. For traditional machine learning methods, Koelstra et al. [10] adopt Gaussian naive Bayes to classify the degree of arousal, valence and liking. Bahari et al. [11] use recurrence plot analysis and k-nearest neighbor classifier to recognize different emotions. Wang et al. [12] employ a support vector machine (SVM) classifier to discriminate different emotions. 

Compared with traditional methods, deep learning technologies have advantages in high-level representation and training in an end-to-end mode. Inspired by the achievements of deep learning in computer vision, natural language processing and other fields [13,~14,~15], deep learning methods are applied to EEG emotion classification. Zheng et.al [16] introduce deep belief networks to investigate critical frequency bands and channels, which is a pioneering work in EEG-Based emotion recognition with neural networks. Since then, many works have utilized deep learning models to exploit EEG data from the properties of spatial, spectral and temporal dimensions. Al-Nafjan et al. [17] adopt Deep Neural Networks (DNN) using PSD feature to identify human emotions. Yang et al. [18] propose a hierarchical network using ~ DE features from five frequency bands to identify different emotions. Above works mainly utilize frequency features but less explore the spatial and temporal information.

To explore the temporal information,  Fourati et al. present an Echo State Network (ESN), which uses recursive layer projects the raw EEG
signals to the high-dimensional state space [19]. Alhagry et al. [20] use a two-layer long-short term memory (LSTM) to get satisfactory emotion recognition results with the EEG signal as input. Ryan et al. [21] use a recurrent neural networks (RNNs) model to explain the time dependence of cognitive-related EEG signal. Bashivan et al. [22] propose a deep recursive convolutional neural network (R-CNN) for EEG-based cognitive and mental load classification tasks. Among diverse deep learning models, Convolutional Neural Network (CNN) [23] is the kernel of the current best architecture for image and video recognizing for their robust capabilities to learn feature representations of input data. CNN is also widely used for spatial features extraction of EEG signal, Li et al. [24] present a hierarchical CNN to capture spatial information among different channels based on 2D EEG maps.  Zhang et al. [25] propose a deep CNN model to learn the spatio-temporal robust feature representation of the raw EEG data stream for motion intention classification.  Lawhern et al. [26] propose an application of multi-layer pure CNN without full connection layer for P300-based oddball recognition task, finger motor task and motor imagination task.  In order to capture the information of interaction between different brain regions, Salama  et al. [27] use a 3D-CNN to recognize human emotions from multichannel EEG data of the DEAP dataset. 

However, CNN is capable of catching local visual field information but not the global information. To overcome the shortcoming, many studies introduce attention mechanism in CNN. Attention can not only learn global information in CNN, but also enhance important information. Even in other deep learning models, attention is used to enhance the vital information as well as suppress unnecessary noise information so as to improve the performance of models. Chen et al. [28] propose a hierarchical bidirectional Gated Recurrent Unit (GRU) network with attention for human emotion classification from continual EEG signal. Kim, et al. [29] propose a long short-term memory network and apply an attention mechanism to assign weights to the emotional states appearing at specific moments to conduct two-level (low and high) and three-level (low, middle, and high) classification on the valence and arousal emotion models. Tao et al. [30] propose an 
 attention-based convolutional recurrent neural network (ACRNN) to extract more discriminative features from EEG signal and improve the accuracy of emotion recognition on the DEAP and DREAMER databases.  Liu et al. [31]  propose a novel multichannel EEG emotion recognition method based on sparse graphic attention long short-term memory (SGA-LSTM) for EEG emotion recognition. Though existing emotion recognition methods have achieved promising results, most methods consider single feature or a combination of two features. Due to the complementarity among features of different dimensions and the different importance of features in different positions of the same dimension to emotion classification, we propose the PST-Attention, which takes into account spatial-spectral-temporal features and the most important features among them simultaneously.

\section{METHODOLOGY}

\begin{figure*}{}
\centering
\includegraphics [scale=0.5]{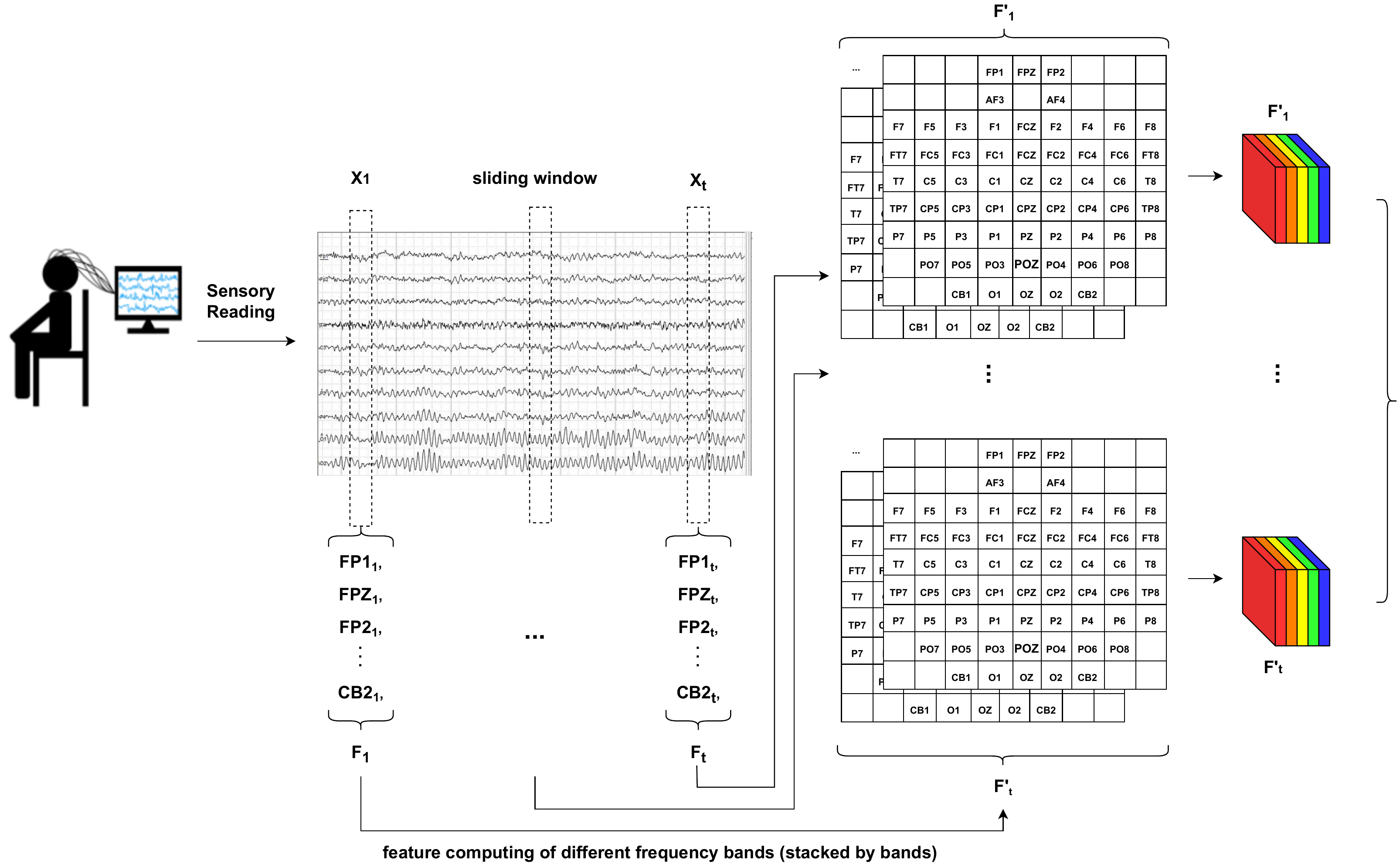}
\centering
\caption{4D EEG Representation Organization }
\label{fig: feature_representation}
\centering
\end{figure*}
\begin{figure}[ht]
\centering
\includegraphics[scale=0.7]{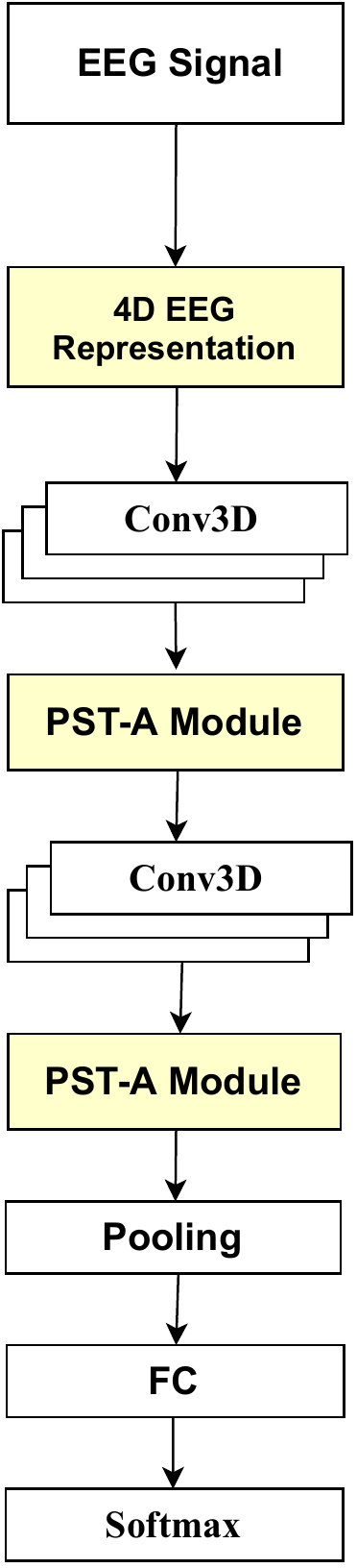}
\caption{Overview of EEG-based emotion recognition.}
\label{fig:overview}
\end{figure}
\begin{figure}[]
  \flushleft 
  \includegraphics [scale=0.63]{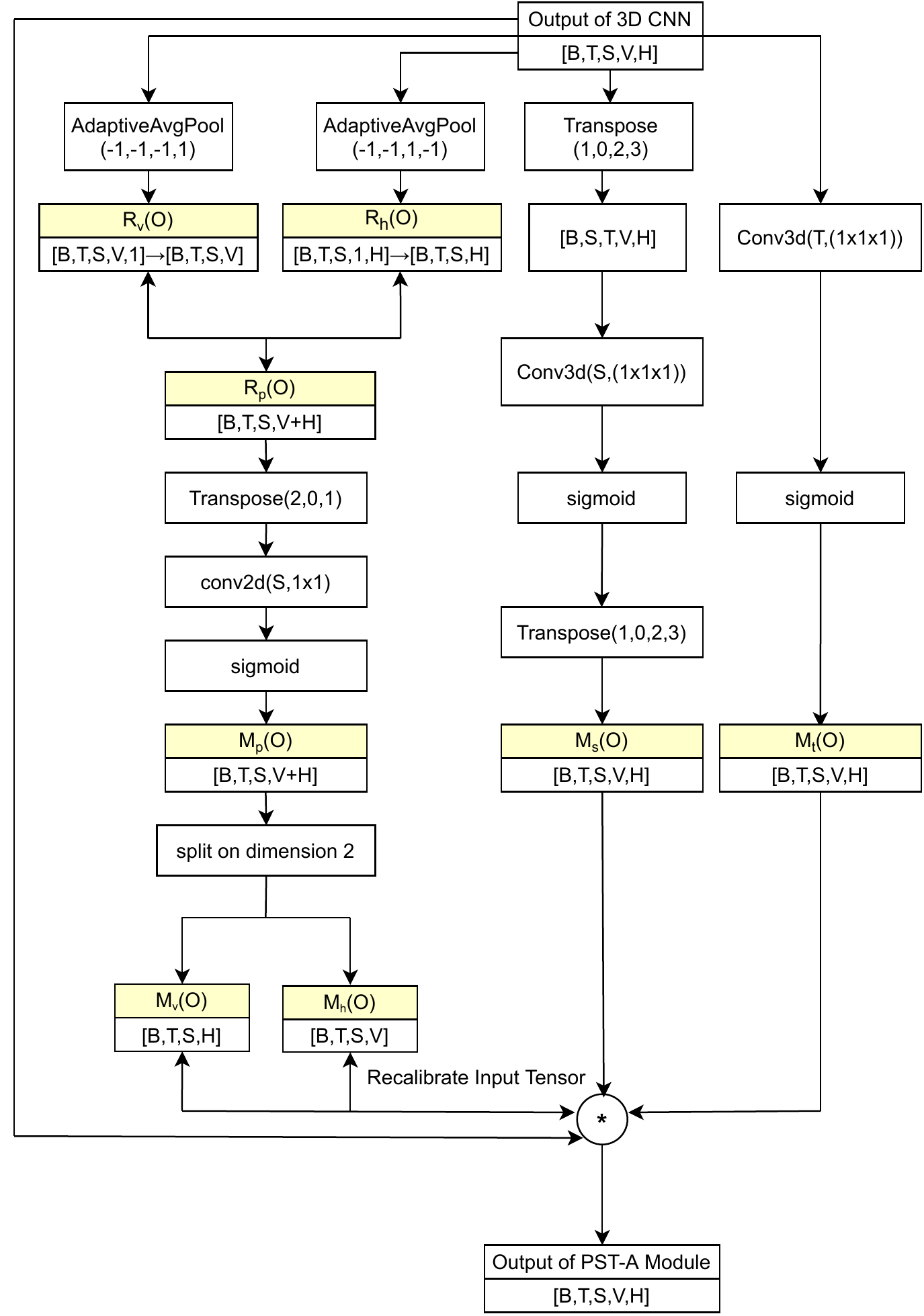}
  \caption{Details of the PST-Attention Module.}
  \label{fig:details}
\vspace{-0.1cm}
\end{figure}
\subsection{Overview of Model Structure}
EEG-based emotion recognition is to classify the emotion categories according to the EEG signal. As illustrated in Fig.~\ref{fig:overview}, the overview of the recognition model in this paper consists of three sub-modules which are feature preparing and re-organization, the backbone of model and classification loss. The yellow parts in Fig.~\ref{fig:overview} are our contributions in this paper. The EEG emotion classification problem to learn a mapping function $F$ that maps the EEG input to the  corresponding emotion labels:
\begin{equation}
Y=F(X^{'})
\label{DKL}
\end{equation}
where $X^{'}$ denotes the representation of EEG signals, $F$ denotes the mapping function i.e., neural network transformations. $Y \in \{y_1, y_2, ... , y_n\}$ denotes the emotion classification labels. In this paper, the classification cross entropy is adopted as loss function, which is defined as the formula:
\begin{equation}
L=-\sum_{c=1}^{C}y_{c}log(y^{'}_{c})
\label{DKL}
\end{equation}
where $L$ denotes the loss function of the task of EEG emotion recoginition, $C$ denotes number of emotion classes, $y_{c}$ is the ground truth emotion label and $y^{'}_{c}$ is the predictors of neural networks. 

The innovation of our paper is the PST-Attention combined with 3D convolution operation (Conv3D) to extract informative feature from the 4D EEG representation. 3D-CNN can better capture the temporal and spatial feature information in EEG signals. The pooling layer in the model is to connect the convolution and full connected layer. The details of EEG representation organization and the PST-Attention module are illustrated in Fig.~\ref{fig: feature_representation} and Fig.~\ref{fig:details} respectively. 

\subsection{4D EEG Representation Organization}
To better represent the features of different dimensions of EEG data, we re-organize EEG signal into a 4D representation. The processing of transforming the original EEG signal into 4D Representation is shown in Fig. \ref{fig: feature_representation}. We define $X=(E_{1},E_{2},...,E_{T})\in \mathbb{R}^{C\times T}$ as an EEG sample collected in $T$ time stamps, where $C$ is the number of electrodes. $E_{t}$ denotes the EEG signal of $C$ electrodes collected at time stamp $t$. Here, we 
use the DE features denoted as $F_{t}=(D_{1},D_{2},...,D_{S})\in \mathbb{R}^{C\times S}$ from $E_{t}$ as described in [32]. We set ($\delta$[1-4Hz], $\theta$ [4-8Hz], $\alpha$  [8-14HZ], $\beta$ [14-31Hz], and $\gamma$ [31-51 Hz]) as the spectral band set $S$. To explore the interactions among spatial, spectral and temporal dimensions, we re-organize $F_{t}$ of the sample $X$ into 4D EEG representation. Specifically, the $s$th band feature  $D_{s}$ from $C$ channels is transformed into a 2D map $D_{s}^{'}\in \mathbb{R}^{V\times H}$ according to the EEG cap layout for EEG channels as Fig.~\ref{fig: feature_representation}.  In other words, we reshape the 1D tensor $D_s \in R_{C}$ into 2D tensor $D_{s}^{'}\in \mathbb{R}^{V\times H}$ $(C \leqslant (V \times H))$. When we do the same operation in each band, $F_{t}$ will be transformed into a 3D map $F_{t}^{'}=(D_{1}^{'},D_{2}^{'},...D_{S}^{'})\in R^{S\times V\times H}$. Finally we stack all the transformed 3D feature map along the temporal dimension to get the 4D EEG representation $X^{'}=(F_{1}^{'},F_{2}^{'},...,F_{T}^{'})  \in \mathbb{R}^{T \times S\times V\times H}$. It is also the form of input for our proposed model. 

\subsection{Proposed PST-Attention Module}
Fig.~\ref{fig:details} illustrates the overview of the Positional-Spectral-Temporal Attention~(PST-Attention).
It consists of a Positional Attention module, a Spectral Attention module and a Temporal Attention module.

\subsubsection{Positional Attention module~(P-Attention)}

The Positional Attention module is to learn the spatial attention mask to enhance the valuable regions of EEG signal for emotion classification. The researches [8] have proved that the activation of different brain regions is dissimilar when subjects watching various videos to evoke emotion. Because there is evidence that the lateralization between the left and right hemispheres is associated with emotions [33], we investigate asymmetry features. Unlike previous spatial attention [30], which learns a 2D attention map within the spatial dimension, we concatenate the relationship on all sides to get global information in the spatial dimensions. Our positional attention is illustrated in the left branch in Fig.~\ref{fig:details}. 
Given the input tensor $X \in \mathbb{R}^{B \times T \times S \times V \times H}$ of the PST-Attention module and $B$ is the batch size in training, the positional attention is formulated as follows:
\begin{equation}
\begin{aligned}
    R_h (X) = \frac{1}{H} \sum_{i=1}^{H}X_{i},
\label{PA_H}
\end{aligned}
\end{equation}

\begin{equation}
\begin{aligned}
   R_v (X) = \frac{1}{V} \sum_{i=1}^{V}X_{i},
\label{PA_W}
\end{aligned}
\end{equation}
\begin{equation}
\begin{aligned}
    M_{vh} = \theta (f_{2d}(\tau([R_v(X),R_h(X))])),
\label{PA_WH}
\end{aligned}
\end{equation}
where $R_v (X)$ and $R_h (X)$ is adaptive average pooling method in horizon and vertical, respectively. $[\cdot~,~\cdot]$ is to concentrate the $R_v(X)$ and $R_h(W)$ alongside temporal dimension. $\tau$ is transpose the tensor $[R_v(X),R_h(X)]$ into $X^{((V+H) \times T \times S)}$. $f_{2d}$ is the attention learning method with 2D convolution operation.  The learning parameters of $f_{2d}$ is~$W^{(V+H)}$ which capture the global context information in horizon and vertical dimensions. Then with the sigmoid function $\theta$ normalizing the attention weight $W^{(V+H)}$, the attention mask $M_{vh} \in \mathbb{R}^{(B,T,S,V+H)}$  is learned. Finally we split  $M_{vh}$ into $M_{v} \in {\mathbb{R}^{(B \times T \times S \times 1 \times H)}}$   and $M_{h} \in \mathbb{R}^{(B \times T \times S \times V \times 1)} $. 

\subsubsection{Spectral Attention module~(S-Attention)}

The Spectral Attention module is to learn the spectrum attention mask to strengthen the informative spectral bands for emotion classification. When the brain is faced with different emotional stimuli, the energy spectrum of different bands will vary greatly. The Spectral Attention module aims to learn the most discriminative EEG frequency that can enhance the ability of emotion classification. The Spectral Attention is illurstrated in the middle branch in Fig.~\ref{fig:details}. The formula is as following:
\begin{equation}
\begin{aligned}
     M_{c} =  \theta (f'_{3d}(\tau(X))))
\label{PA_T}
\end{aligned}
\end{equation}
where $\tau$ is the transpose operation to exchange the temporal and spectral dimensions. $f'_{3d}$ is a 3D convolution operation with kernel size $(S \times 1 \times 1)$.  The $W^{S \times 1 \times 1}$ is the learning weight represent the different importance of spectral-wise. $\theta$ is the sigmoid function to scale the spectral-wise attention weight $W^{S}$. 

\subsubsection{Temporal Attention module~(T-Attention)}

The Temporal Attention module is to learn the time dimension mask to reinforce the climax time slice of a particular emotion. The generation of emotions can't accomplish at one stroke, it may experience upsurge or downcast. So for the same emotion at different time points, the situation of brain waves is also different. We take advantage of the Temporal Attention module to learn changes of a certain emotion in the time axis. The Temporal Attention is illustrated in the right branch in Fig.~\ref{fig:details}. The Temporal Attention formula is as following:
\begin{equation}
\setlength\abovedisplayskip{0.1cm}
\setlength\belowdisplayskip{0.1cm}
\begin{aligned}
     M_{t} =  \theta (f''_{3d}(X)),
\label{PA_T}
\end{aligned}
\end{equation}
where the $f''_{3d}$ is the convolution operation with kernel size $( T \times 1 \times 1)$. The parameters weight of $f''_{3d}$ is $W^{T}$, which represent the difference importance of different temporal. $\theta$ is sigmoid function.

The $M_{vh}$, $M_s$ and $M_t$ are Positional Attention mask, Spectral Attention mask and Temporal Attention mask respectively. We call the whole attention module as Positional-Spectral-Temporal Attention~(PST-Attention).  After attention masks learning, the input tensor is recalibrated by the three attention masks. In fig.~\ref{fig:details}, the $*$ means the dot multiplication operation.

\section{Experiments}
\subsection{Datasets}
We validated our model on SEED [8,~32] and SEED-IV [34] databases.

\textbf{SEED} contains three different categories of emotion, namely positive, negative, and neutral. Fifteen participants’ EEG data of the dataset were collected while they were watching the stimulus videos. The videos are carefully selected and can elicit a single desired target emotion. With an interval of about one week, each subject participated in three experiments, and each session contained 15 film clips.
The participants are asked to give feedback immediately after each experiment. The EEG signals of 62 channels are recorded at a sampling frequency of 1000 Hz and down-sampled with 200 Hz. The DE features are pre-computed over different frequency bands for each sample in each channel.

\textbf{SEED-IV} The dataset contains four different categories of emotions, including happy, sad, fear, and neutral emotion. The experiment consists of 15 participants. Three experiments are designed for each participant on different days, and each session contains 24 video clips and six clips for each type of emotion in each session.  After each experiment, the subjects are asked to give feedback, while 62 EEG signals of the subjects are recorded. The EEG signals are sliced into 4-second non-overlapping segments and down-sampled with 128 Hz. The DE feature is also pre-computed over five frequency bands in each channel.

\subsection{Experimental Setup}
We train our model on NVIDIA RTX 2080 GPU. The Adam optimization is used to minimize the loss function. The learning rate is set to 0.001. The kernel size of the 3D-CNN is 5*5*3. The number of the Attention Module Block is set to 2 and the number of time stamps in each sample is set to 9, considering the longer temporal feature may have more meaningful Physiological significance, different from [41], which sets the time slice length to 0.5s, we set the time slice length to 1s. We conducted experiments on each subject. For each experiment, we randomly shuffle the samples. The ratio of the training set to test set is 9:6.

\subsection{Compared Models}
• SVM [35]: A Least squares support vector machine classifier.

• DBN [16]: Deep Belief Networks investigate the critical frequency bands and channels.

• DGCNN [36]: Dynamical Graph Convolutional Neural Networks model the multichannel EEG features.

• GSCCA [37]: Group sparse canonical correlation analysis, a group sparse extension of the conventional CCA method which models the linear correlationship between emotional EEG feature vectors and the corresponding EEG class label vectors.

• BiDANN [38]: Bi-hemispheres domain adversarial neural network maps the EEG feature data of both left and right hemispheres into discriminative feature spaces separately.

• RGNN [39]: Regularized graph neural network considers the biological topology among different brain regions to capture both local and global relations among different EEG channels.

• BiHDM [40]: Bi-hemispheric discrepancy model learns the asymmetric differences between two hemispheres for EEG
emotion recognition.

• 4D-aNN [41]: Four-dimensional attention-based neural 
network fuses information on different domains 
and captures discriminative patterns in EEG signals based on 
the 4D spatial-spectral-temporal representation.

\subsection{Experimental Results and Analysis}
We compare our model with other models on SEED and SEED-IV
dataset. TABLE \ref{results} presents the average accuracy (Mean) and standard deviation (Std) of the compared models for EEG based emotion recognition on SEED and SEED-IV datasets.  DGCNN [36] considers the spatial information of EEG signals collected from different channels and employs graph convolution to extract spatial information. BiHDM [40] uses two directional RNNs to extract spatial information of EEG signals. 4D-aNN [41] transforms DE and PSD features into 4D spatial-spectral-temporal representations as input and integrates the attention mechanisms into the CNN module and the bidirectional LSTM module for emotion classification. Our model takes positional, spectral, and temporal information into consideration simultaneously. In light of the asymmetry of brain regions that significantly influence emotion classification, we design the Positional Attention to concatenate the relationship on all sides to get global information in the spatial dimensions. It learns the critical information of anterior-posterior and left-right hemispheres simultaneously, enabling our model to capture informative features from EEG signals for emotion recognition comprehensively. In addition, with the Spectral Attention module to learn critical frequency bands and the Temporal Attention module to learn important time points, the accuracy of our model is further improved compared with the other competitive models. 

The TABLE \ref{results} presents the performance of all models on the SEED and the SEED-IV dataset. For the three-category classification task, 4D-aNN [41] transforms the combination of DE and PSD features to input, reaching 96.10$\%$ on classification accuracy. Xiao et al. [41] also conduct experiments by only taking DE features and PSD features as inputs, respectively. The 4D-aNN (DE) accuracy reaches 95.39$\%$ and exceeds 4D-aNN (PSD) 4.90$\%$, and the result is very close to 4D-aNN’s experiments using both of them. The performance of our method is better than that of 4D-aNN(DE), indicating our effectiveness. For the four-category classification task, DBN achieved an accuracy of 66.77$\%$, and the graph-based network DGCNN [36] and RGNN [39] further improve the accuracy to 69.88$\%$ and 79.37$\%$, respectively. 3D-CNN with PST-Attention achieves superior performance with an accuracy of 82.73$\%$ on SEED-IV datasets compared to other competitive models.

To verify the validity of 4D feature representation, we conduct ablation experiments with different EEG sample representations on SEED and SEED-IV datasets. TABLE \ref{abc} presents the experimental results on different EEG representations, where VHS stands for  spatial (Horizontal and Vertical) and spectral dimensions. VHT stands for the spectral and temporal dimensions, as for spectral dimension, we take the best result from the feature of 5 different frequency bands and the mean value. PST means using spatial, spectral, and temporal dimensions without split the spatial dimension into Horizontal and Vertical dimensions. The results show that the 4D representation obtains better accuracy in terms of the same structure than other competitive EEG representations. As for the 3D EEG representation, we use a more suitable structure for 3D input, CNN, to conduct emotion recognition \ref{abc}. 

\begin{table}[]
\centering
\caption{The performance comparison of the competitive models on SEED and SEED-IV dataset}
\resizebox{\linewidth}{!}{
\begin{tabular}{cclcl}
\hline
\multirow{2}{*}{\textbf{Models}} & \multicolumn{2}{c}{\textbf{SEED}} & \multicolumn{2}{c}{\textbf{SEED-IV}} \\
                                 & Mean (\%)         & Std (\%)        & Mean (\%)          & Std (\%)          \\ \hline
SVM {[}35{]}                     & 83.99            & 9.72           & 56.61             & 20.05            \\
GSCCA {[}37{]}                   & 82.96            & 9.95           & 69.08             & 16.66            \\
DBN {[}16{]}                     & 86.08            & 8.34           & 66.77             & \textbf{7.38}             \\
DGCNN {[}36{]}                   & 90.40            & 8.49           & 69.88             & 16.29            \\
BiDANN {[}38{]}                  & 92.38            & 7.04           & 70.29             & 12.63            \\
RGNN {[}39{]}                    & 94.24            & 5.95           & 79.37             & 10.54            \\
BiHDM {[}40{]}                   & 93.12            & 6.06           & 74.35             & 14.09            \\
4D{-}aNN{(}DE{)} {[}41{]}                   & 95.39             &\textbf{3.05}           & {-}                & {     -}            \\

\textbf{3D-CNN \& PST-Attention}                         & \textbf{95.76}            & 4.98           & \textbf{82.73}             & 8.96             \\ \hline
\end{tabular}
}
\label{results}
\end{table}

\begin{table}[]
\centering
\caption{ The Experimental Results of the 4D Representation on SEED and SEED-IV dataset}
\begin{tabular}{cccccc}
\hline
\multirow{2}{*}{\textbf{Representation}} & \multicolumn{1}{l}{\multirow{2}{*}{\textbf{Structure}}} & \multicolumn{2}{c}{\textbf{SEED}}                     & \multicolumn{2}{c}{\textbf{SEED-IV}}                  \\
                                         & \multicolumn{1}{l}{}                                    & \textbf{Mean}             & \textbf{Std}              & \textbf{Mean}             & \textbf{Std}              \\ \hline
3D (VHS)                                 & CNN                                                     & 85.43                     & 17.68                     & 68.45                     & 18.97                     \\
3D (VHT)                                 & CNN                                                     & 86.31                     & 13.91                     & 65.86                     & 17.63                     \\
3D (PST)                                 & CNN                                                     & \multicolumn{1}{l}{83.63} & \multicolumn{1}{l}{15.36} & \multicolumn{1}{l}{66.37} & \multicolumn{1}{l}{17.37} \\
\textbf{4D (VHST)}                       & 3D-CNN                                                  & \textbf{92.79}            & \textbf{11.93}            & \textbf{72.29}            & \textbf{15.76}            \\ \hline
\label{abc}
\end{tabular}
\end{table}

\begin{figure}[]
\centering
\includegraphics[scale=0.3]{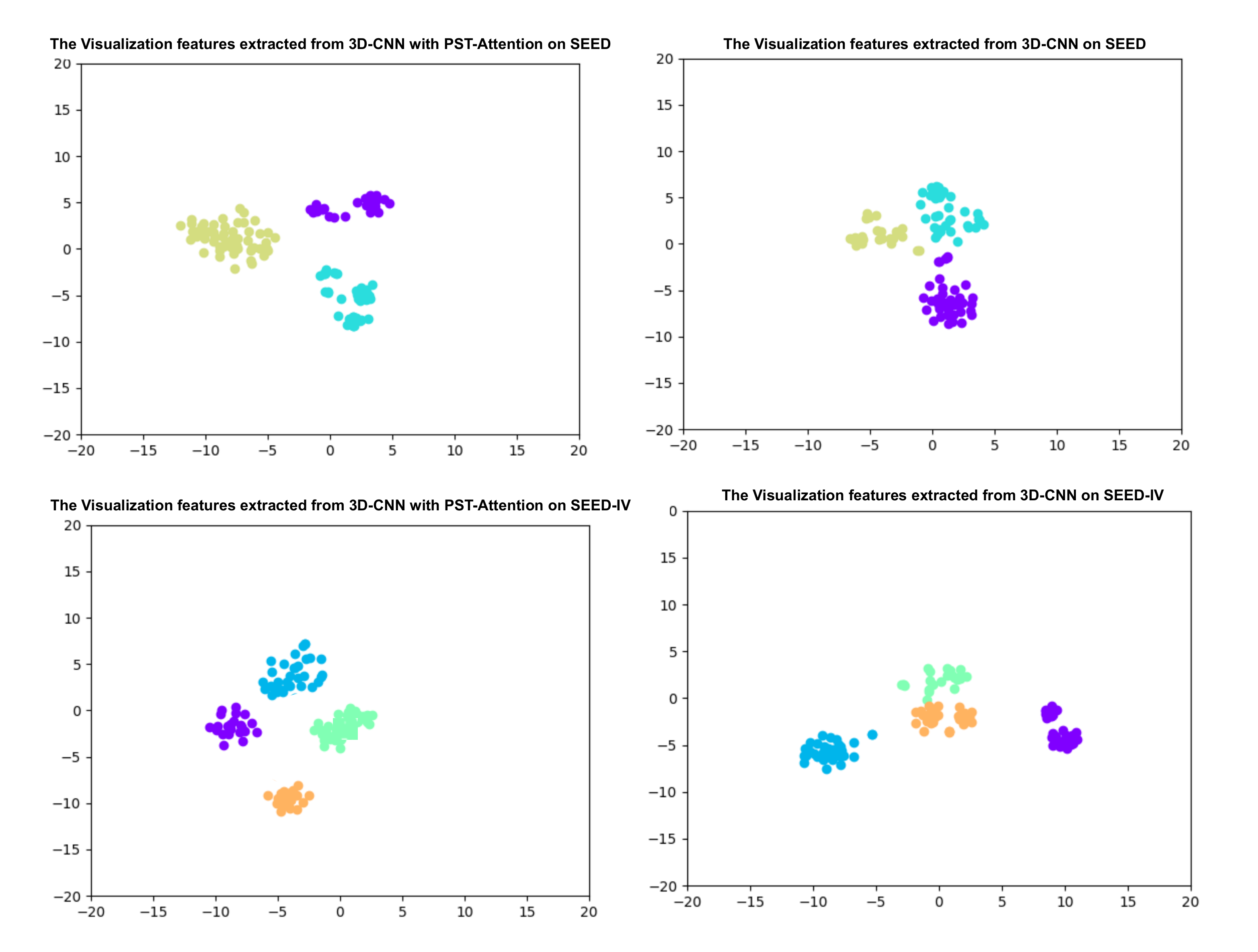}
\caption{Visualization of the high-level features extracted from the PST-Attention in 3D-CNN.}
\label{fig:analysis}
\end{figure}

To further validate the performance of the proposed method, we visualize the high-level bottleneck features from well trained model to analyze. We extract the bottleneck features after the pooling layer from the well trained model, and visualize the features our test datasets of SEED and SEED-IV in Fig.~\ref{fig:analysis}. Comparing these four pictures, the distances of different classes in the high-level feature space of 3D-CNN with PST-Attention are more dispersed than that of 3D-CNN. This demonstrates that the high-level features learned from 3D-CNN with PST-Attention is more discriminative than that learned from 3D-CNN.

\subsection{Ablation Studies}
To demonstrate the effectiveness of different modules in PST-Attention, we conduct ablation experiments on SEED-IV database. The proposed PST-Attention module includes three submodules, which are P-Attention, S-Attention, and T-Attention. TABLE \ref{ab} presents the experimental results on each sub-modules. The results of the three modules have different degrees of improvement in classification accuracy. In TABLE \ref{ab}, 3D-CNN with P-Attention achieves 3.27\% improvements on Mean and 3.02\% reductions on Std. 3D-CNN with S-Attention achieves 2.47\% improvements on Mean and 3.84\% reductions on Std. 3D-CNN with T-Attention achieves 0.4\% improvements on Mean and 1.8\% reductions on Std. Grouping these three attention modules into 3D-CNN achieves 9.81\% improvements on Mean and 6.8\% reductions on Std.  These illustrate that different attention modules possess complementary in mining useful EEG information for emotion recognition. We also conclude that the effectiveness of the P-Attention or S-Attention module is a bit better than that of the T-Attention module, which indicates that the informative channels and frequency bands may have more importance on emotion recognition. The Positional Attention mask and Spectral Attention mask show that the features of different brain regions and different frequency bands play different roles in EEG emotion recognition, consistent with the literature observations [6,~8].
\begin{table}[tbp]
\caption{ The Ablation Experimental Results of the Proposed PST-Attention on SEED-IV dataset}
\centering
\begin{tabular}{ccc}
\hline
\textbf{Structures}                                                                  & \multicolumn{1}{l}{Mean (\%)} & \multicolumn{1}{l}{Std (\%)} \\ \hline
3D-CNN                                                                               & 72.29                        & 15.76                       \\
\begin{tabular}[c]{@{}c@{}}3D-CNN \& P-Attention\end{tabular}           & 75.56                        & 12.74                       \\
\begin{tabular}[c]{@{}c@{}}3D-CNN \& S-Attention \end{tabular}             & 74.76                        & 11.92                        \\
\begin{tabular}[c]{@{}c@{}}3D-CNN \& T-Attention \end{tabular}             & 72.69                        & 13.96                       \\
\begin{tabular}[c]{@{}c@{}}\textbf{ 3D-CNN \& PST-Attention}      \end{tabular} & \textbf{82.73}                       & \textbf{8.96}                       \\ \hline
\label{ab}
\end{tabular}
\end{table}

\section{CONCLUSION}
In this paper, we propose the PST-Attention in 3D-CNN for EEG emotion recognition. The 3D-CNN with PST-Attention model extracts positional features, spectral features, and temporal features with the proposed  4D EEG representation, which effectively utilizes the complementarity among features of different dimensions, enhances the informative features as well as suppress useless features. In addition, the PST-Attention mechanism is used to adaptively focus on attentive brain regions, brain asymmetry, frequency bands, time stamps. The experiments on SEED and SEED-IV datasets demonstrate better performance than all competitive methods. Meanwhile, ablation studies show the effectiveness of each attention module in the PST-Attention. The 3D-CNN with the proposed PST-Attention is a general framework based on multivariate physiological time series, which can be further applied to other fields in the future, such as pressure detection and sleep quality analysis.

\end{document}